\documentclass[aps,prd,amsmath,amssymb,preprintnumbers,onecolumn,12pt,nofootinbib]{revtex4}

\usepackage{mathpazo}
\usepackage[utf8]{inputenc}
\usepackage[english]{babel}
\usepackage{amsmath,amssymb}
\usepackage{mathrsfs}
\usepackage{graphicx}
\usepackage{enumerate}
\usepackage{slashed}

\newcommand{\threetwo}{\langle \sigma v^2 \rangle_{3\to 2}}

\newcommand{\cO}{\mathcal{O}}
\newcommand{\beq}{\begin{equation}}
\newcommand{\eeq}{\end{equation}}
\newcommand{\bea}{\begin{eqnarray}}
\newcommand{\eea}{\end{eqnarray}}

\begin{document}

%%%%%%%%%%%%%%%%%%%%%%%%%%%%%%%%%%%%%%%%%%%%%%%%%%%%%%%%%%%%%%%%%%%%%%%%%%%%%%%%%%%%%%%%
\title{\Large
Hidden strongly interacting massive particles
}
%%%%%%%%%%%%%%%%%%%%%%%%%%%%%%%%%%%%%%%%%%%%%%%%%%%%%%%%%%%%%%%%%%%%%%%%%%%%%%%%%%%%%%%%

\author{Matti {\sc Heikinheimo}}
\email{matti.heikinheimo@helsinki.fi}
\affiliation{Department of Physics, University of Helsinki
\& Helsinki Institute of Physics, \\
                      P.O.Box 64, FI-00014 University of Helsinki, Finland}

\author{Kasper {\sc Lang\ae ble}}
\email{langaeble@cp3-origins.net}
\affiliation{CP$^3$-Origins, University of Southern Denmark,
Campusvej 55, 5230 Odense M, Denmark}

\author{Kimmo {\sc Tuominen}}
\email{kimmo.i.tuominen@helsinki.fi}
\affiliation{Department of Physics, University of Helsinki,
\& Helsinki Institute of Physics, \\
                      P.O.Box 64, FI-00014 University of Helsinki, Finland.}

\begin{abstract}
	We consider dark matter as Strongly Interacting Massive Particles (SIMPs)
	in a hidden sector, thermally decoupled from the Standard Model heat bath.
	Due to its strong interactions, the number-changing processes of the SIMP
	lead to its thermalization at temperature $T_{\rm{D}}$ 
	different from the visible sector temperature $T$, and subsequent
	decoupling as the Universe expands. We study the evolution of the
	dark SIMP abundance in detail and find that a hidden SIMP provides for
	a consistent framework for self-interacting dark matter. Thermalization and decoupling of a composite SIMP can be treated within the domain of validity of chiral perturbation theory unlike the simplest realizations of the SIMP, where the SIMP is in thermal equilibrium with the Standard Model.\\[5mm]
	{\it Preprint: HIP-2018-9/TH \& CP3-Origins-2018-010 DNRF90}
\end{abstract}

\maketitle

\section{Introduction}\label{sec:intro}

Dark matter (DM) constitutes most of the matter in the Universe~\cite{Ade:2015xua}. Yet, besides
its overall abundance, only very little is known about its nature.
The dark matter problem has provided significant impetus for construction
of beyond the Standard Model (SM) theories,
and currently there exists many paradigms for dark matter.

In a wide class of models~\cite{Baer:2014eja} the relic abundance is generated via a thermal freeze-out, where typically a $2\rightarrow 2$ process keeps dark matter
in thermal equilibrium with the SM particles until the rate of the dark matter annihilation
process drops below the Hubble rate and dark matter freezes out.
This requires a sufficiently large coupling to deplete the dark matter abundance to the observed value. As a result, the WIMP models are expected to be testable in direct and indirect detection, as well as in collider experiments. As no convincing signals from these searches have emerged~\cite{Akerib:2016vxi,Cui:2017nnn,Ackermann:2015zua,Slatyer:2015jla}, the standard WIMP scenario is beginning to look less convincing. Very recently, even tighter constraints for DM annihilation into SM states have been obtained~\cite{DAmico:2018sxd} from the 21 cm absorption signal observed by the EDGES experiment~\cite{Bowman:2018yin}.

The constraints that have ruled out a majority of the natural parameter space of the standard WIMP scenario have motivated an increasing focus on a different paradigm: If the coupling of the dark matter to the SM fields is very feeble, it is
still possible to produce the observed dark matter density out of
equilibrium by thermal scattering from the SM fields into hidden sector states \cite{McDonald:2001vt,Choi:2005vq,Petraki:2007gq,Hall:2009bx}.
For a review of the recent progress in the freeze-in paradigm, see~\cite{Bernal:2017kxu}.

Yet another possibility for producing the DM abundance is provided by number-changing processes, such as $3\rightarrow 2$, involving just the DM which in this case consists of strongly interacting
massive particles (SIMP). This process reduces the number of dark matter
particles while simultaneously heating them up~\cite{Carlson:1992fn,deLaix:1995vi,Hochberg:2014dra,Hansen:2015yaa}. This scenario does not depend on DM annihilations to the SM states, and is therefore not constrained by most of the bounds that affect the WIMP scenario. Strongly coupled dynamics is also
appealing due to the possible connections with dark matter self-interactions
which could resolve some discrepancies in the small scale structure formation
like missing satellites and cuspy vs. cored density profiles of galaxies~\cite{Tulin:2017ara}.

This SIMP paradigm, however, suffers from an internal inconsistency:
The leading order (LO) analysis is phenomenologically unreliable as it is outside the range of convergence of
chiral perturbation theory \cite{Hansen:2015yaa}. Moreover, after including
next-to-leading order (NLO) and next-to-next-to-leading order (NNLO) corrections into the
chiral perturbation theory treatment, maintaining the viability of the simplest SIMP realizations\cite{Hochberg:2014kqa} in light of the phenomenological constraints becomes difficult~\cite{Hansen:2015yaa}. Outside the applicable range of chiral perturbation theory, explicit inclusion of resonances in the Lagrangian become necessary, and this potentially has a dramatic effect on the predicted mass range of the composite SIMP \cite{Berlin:2018tvf,Choi:2018iit}.

In this paper we extend the analysis to another direction, by relaxing the assumption of kinetic equilibrium between the hidden sector and the SM heat bath during the DM production phase. Instead, we will consider
a strongly coupled sector only feebly coupled with the SM and therefore
not in thermal equilibrium with the SM in the early Universe.
After initial population of the hidden sector,
the $2\leftrightarrow 3$ processes will bring  it into internal thermal equilibrium
at $T_{\rm{D}}\neq T$. Eventually, as the Universe expands,
these processes are no longer able to maintain chemical equilibrium within the hidden sector. The SIMP freezes out as the scattering rate of the $3\rightarrow 2$ process drops below the Hubble rate.

We will show that this framework leads to a viable dark matter candidate. Furthermore, we will show that our analysis can be consistently carried out
within the range of convergence of the chiral perturbation expansion and that it
increases the viable parameter space of the simplest SIMP models.

The paper is organized as follows.
In Sec.~\ref{sec:StandardSIMP}, we briefly review the SIMP
mechanism~\cite{Carlson:1992fn,deLaix:1995vi,Hochberg:2014dra} and its simplest
realization~\cite{Hochberg:2014kqa}. In these cases, the SIMPs are kept in
kinetic equilibrium with the visible sector such that at freeze-out $T_{\rm{D}} =
T$. Then, in Sec.~\ref{sec:DarkFreezeOut}, we relax the requirement of
kinetic equilibrium between the two sectors and derive an estimate for the
magnitude of the $3\to2$ cross section to produce the observed relic
abundance of dark matter.
By specializing to the concrete realization, we illustrate directly how the predicted
DM mass range and perturbativity of the model depend on the temperature ratio
of the sectors at the time of freeze-out. We discuss the possible origins of
the two sectors, constraints on this scenario and means to test the model in
Sec.~\ref{sec:Discussion}. Finally, we conclude in
Sec.~\ref{sec:Conclusion}.

\section{The standard SIMP}\label{sec:StandardSIMP}
The SIMP scenario provides an alternative mechanism to thermally produce the
observed DM relic density. Instead of using  $2\to2$ annihilation processes,
one assumes that a dominant $3\to2$ number-changing process involving only the
SIMPs occurs in the dark sector. This process reduces the number of dark
particles at the cost of heating up the sector. Despite this, the DM particles
remain in kinetic equilibrium with the standard model photons if a small
coupling between the dark and visible sectors is assumed.
In this way the energy from the dark sector can be transferred to the
SM sector via elastic scattering processes.

The cross sections of the $3\rightarrow 2$ and the elastic self-scattering processes in the model are parametrized as
\begin{equation}
\label{SIMP-coupling}
\langle\sigma v^2\rangle_{3\to2}= \frac{\alpha_{\text{eff}}^3}{m_{\text{D}}^5}, \qquad \qquad \frac{\sigma_{\text{scatter}}}{m_{\text{D}}}= \frac{a^2\alpha_{\text{eff}}^2}{m_{\text{D}}^3}\, ,
\end{equation}
where $a\equiv \alpha_{2\to 2}/\alpha_{\text{eff}}$ and is expected to be of order unity.

In the case where the $3\to2$ process dominates over the $2\to2$ process, the correct relic abundance is produced with the cross section
\begin{equation}\label{FreezeSigma}
\threetwo \simeq 8.65 \,\,  \text{GeV}^{-5}\, x_{\mathrm{FO}}^4\, g_{\mathrm{eff}}^{-3/2} \left(\frac{1\, \text{GeV}}{m_{\text{D}}}\right)^{2}\, ,
\end{equation}
where $x_{\rm{FO}} = m_{\rm{D}} / T$ and $g_{\mathrm{eff}}$ is the number of
effective degrees of freedom. However, subjecting the DM self-interactions to the
observational constraints, roughly $\sigma_{\rm scatter}/m_{\rm D} 
\lesssim 1\ {\rm cm}^2/{\rm g}$, implies $a\sim\mathcal{O}(10^{-1})$ for 
$\alpha_{\rm eff}\sim 1$, or $a\sim\mathcal{O}(1)$ for $\alpha_{\rm eff} \gg 1$. This suggests that unless a specific
realization of the SIMP mechanism provides a suppression for $a$ of order
$\cO(10^{-1})$, the scenario will only be viable for high values of
$\alpha_{\text{eff}}$ and for masses around the GeV scale.~\footnote{In
\cite{Hochberg:2014dra} it is argued that the effective coupling can be
significantly larger than unity if, for example, the number of DM degrees of
freedom is large, if the cross section is nonperturbatively enhanced, or if
the $3\to 2$ process is mediated by a light particle.}

A concrete model building framework of the SIMP mechanism is provided
by a strongly coupled gauge theory described at low energies via
chiral perturbation theory.
The SIMP mechanism was originally realized in such a
setting~\cite{Hochberg:2014kqa}. The model proposed
pions as DM particles, while the number-changing interaction was
the Wess-Zumino-Witten (WZW)
term~\cite{Wess:1971yu,Witten:1983tw,Witten:1983tx}.

Generally, the relevant chiral symmetry breaking pattern
is dictated by the number of colors, fermion flavors and their representation
under the gauge group.
The minimal case is an Sp($N_c$) gauge theory ($N_c$ even) with four Weyl
fermions in the fundamental $N_c$-dimensional representation, which follows the
chiral symmetry breaking pattern $\text{SU}(4)\to\text{Sp}(4)$. The relevant
cross sections can be calculated in chiral perturbation theory, and at lowest nonvanishing order
\begin{equation}\label{eq:crossSIMPlest}
\langle\sigma v^2\rangle_{3\to2} = \frac{5 \sqrt{5} N_c^2 m_{\pi}^5}{2 \pi^5 x^2 f_{\pi}^{10}}\frac{t^2}{N_{\pi}^3}, \qquad \qquad
\frac{\sigma_{\text{scatter}}}{m_{\pi}} = \frac{m_{\pi}}{32\pi f_{\pi}^4} \frac{b^2}{N_{\pi}^2} \, ,
\end{equation}
where $t^2$,  $b^2$ and $N_{\pi}$ depend on the breaking pattern and are given in \cite{Hochberg:2014kqa}. We note that the former is NLO, while the latter is LO in the chiral expansion.

Performing the analysis to the lowest nonvanishing order leads to tension in meeting the observational constraints within perturbation theory for all the minimal cases. The tension is weakened by increasing $N_c$. However, including higher order terms in the chiral expansion indicates that in order for the model to be viable and under perturbative control, we need $N_c \gtrsim 16$~\cite{Hansen:2015yaa}.\footnote{It is pointed out in \cite{Hochberg:2014kqa} that a further explicitly broken flavor symmetry would decrease the value of $b^2$.} The framework has been extended with concrete connections to the SM, and accompanying observables have been studied \cite{Lee:2015gsa,Hochberg:2015vrg,Kamada:2016ois,Kamada:2017tsq,Hochberg:2017khi}. In the next section, we will propose a way for the model to meet observational constraints while being under perturbative control even for the case of $N_c = 2$.

\section{Dark Freeze-out with $T_{\rm{D}} \neq T$}\label{sec:DarkFreezeOut}

In this section we will assume the existence of a strongly interacting
dark sector, which is not in thermal equilibrium with the visible sector (SM),
and study its evolution as the Universe expands.
We will discuss concrete model frameworks leading to such an initial
setup in more detail in Sec.~\ref{sec:Discussion}.

For simplicity we assume that
the hidden sector contains only dark matter in
thermal equilibrium with itself at $T_{\rm{D}}$, while the visible sector has the
temperature $T$. If the DM particles are sufficiently weakly interacting,
the freeze-out happens at a temperature $T_{\rm{D}}\gg m_{\rm{D}}$ when the DM particles are still relativistic. Then, in order to produce the observed DM abundance, the temperature ratio at the time of freeze-out must be
% % %
\begin{equation}\label{relativisticRatio}
\frac{T_{\rm{D}}}{T}=\left(\frac{h_{\mathrm{eff}}}{g_{D}}\frac{2\pi^4\, Y_{\infty}}{45\, \zeta(3)}\right)^{1/3}\, ,
\end{equation}
% % %
where $h_{\mathrm{eff}}$ is the effective number of relativistic degrees of
freedom in the visible sector contributing to the entropy, $g_{D}$ is the
DM degrees of freedom and  $Y_{\infty}$ is the DM yield, \mbox{$m_{\rm{D}}Y_\infty = 4.2\times 10^{-10}\, {\rm GeV}$}, fixed to give the observed relic density.
For comparable numbers of degrees of freedom, the temperature ratio is roughly $10^{-3} \left(\text{GeV}/m_{\rm{D}}\right)^{1/3}$. We can also express this in terms of the ratio of SM entropy to the dark sector entropy, $\xi = S/S_{\rm{D}}$, which defines a mass-dependent upper limit
% % %
\begin{equation}\label{xiRatio}
\xi_0 = \frac{45\, \zeta(3)}{2\pi^4\, Y_{\infty}}.
\end{equation}
% % %
% % %
Instead, if DM interacts more strongly, so that the decoupling happens at a nonrelativistic temperature $T_{\rm{D}} \ll m_{\rm{D}}$, the observed relic density is produced when the temperatures satisfy
% % % %
\begin{equation}\label{xpFO}
x'_{\rm{FO}} = 22 - \ln \left[ \left(\frac{h_{\mathrm{eff}}}{g_{DS}}\right) \left(\frac{x'_{\rm{FO}}}{x_{\rm{FO}}}\right)^{3} \left(\frac{100 \,  \text{MeV}}{m_{\rm{D}}}\right)\left(\frac{x'_{\rm{FO}}}{22}\right)^{-3/2} \right]\, ,
\end{equation}
% % % %
where $x'_{\rm{FO}} = m_{\rm{D}} / T_{{\rm{D}}, {\rm{FO}}}$ and $x_{\rm{FO}} = m_{\rm{D}} / T_{{\rm{FO}}}$. This result can be derived by adopting the standard assumptions to
reduce the problem to a single Boltzmann equation
%% % %
 % %
	%
	\begin{equation}\label{Boltzmann-Y}
	\frac{dY}{dx} = -\sqrt{\frac{4\pi^5}{91125G}}\frac{m_{\rm{D}}^4}{x^5}g_*^{1/2} h_{\mathrm{eff}}\, (Y^3-Y^2Y_{\mathrm{eq}})\langle\sigma v^2\rangle_{3\to2}\,,
	\end{equation}
	where $Y = n(T_{\rm{D}}, m_{\rm{D}}) / s(T)$, $x = m_{\rm{D}} / T$. The factor $g_*$ is the following combination
	\begin{equation}
	g_* ^{1/2} = \frac{h_{\mathrm{eff}}}{\sqrt{ g_{\mathrm{eff}}}}\left(1 + \frac{T}{3 h_{\mathrm{eff}}} \frac{d h_{\mathrm{eff}}}{dT}\right)
	\end{equation}
	where $g_{\mathrm{eff}}$ is the effective number of degrees of freedom contributing to the energy density. Then, assuming the cross section to be independent of the relative velocity \cite{Bernal:2015ova},
  Eqs.~(\ref{FreezeSigma}) and (\ref{xpFO}) follow.

If the two sectors were in kinetic equilibrium during freeze-out, i.e. $T_{\rm{D}} = T$, Eq.~\eqref{xpFO} implies that $x'_{\rm{FO}}\sim 22$ for $g_{DS} \sim h_{\mathrm{eff}}$ and $m_{\rm{D}} \sim 100\,\text{MeV}$.
However, if the two sectors evolve independently at different temperatures, a smaller temperature ratio $T_{\rm{D}}/T < 1$ results in an earlier freeze-out, i.e. $x'_{\rm{FO}}< 22$, whereas a larger ratio corresponds to a later freeze-out. The nonrelativistic assumption breaks down at roughly half the ratio given in Eq.~(\ref{relativisticRatio}), i.e. at $\xi\simeq \xi_0/2$.

We note that Eq.~(\ref{Boltzmann-Y}) depends on $T_{\rm{D}}$ through the equilibrium number density of dark matter and, potentially, through the $3\to2$ cross section.
	
On the other hand, the solution of Eq.~(\ref{xpFO}) for the temperature of the dark sector at freeze-out is only logarithmically sensitive to the temperature ratio of the two sectors. If the $3\to2$ cross section is velocity independent, the value of the cross section that produces the observed abundance, Eq. (\ref{FreezeSigma}), depends solely on the temperature of the visible sector.
Therefore, if the hidden sector is colder than the visible one, the effective
coupling required to produce the correct relic density is reduced.

The above reasoning is based on the assumptions normally applied to the
case where the freeze-out happens via a $2\to2$ process. The fact that the dominating process at freeze-out in our case is $3\to2$, resulting in an extra power of $Y$ in Eq.~(\ref{Boltzmann-Y}) compared to the $2\to2$ case, makes the freeze-out more abrupt. This weakens the assumption $Y(\infty) \ll Y(x'_{\rm{FO}})$. Furthermore, if the $3\to2$ cross section is velocity dependent [as in Eq.~(\ref{eq:crossSIMPlest})] the decoupling is even faster. On top of this, since the dark sector is not in kinetic equilibrium with the standard model photons, the dark sector temperature will, after freeze-out, decrease even faster than the photon temperature.

These considerations suggest another way to approximate the required interaction strength. At sufficiently high temperatures the hidden sector, internally, will be in both kinetic and chemical equilibrium. As the Universe expands, the momentum of the DM particles decreases. However, fast number-changing processes ensure entropy conservation \cite{Carlson:1992fn} and as a result, the dark temperature decreases only logarithmically
	\begin{equation}\label{Tlog}
	T_{\rm{D}} \simeq \frac{m_{\rm{D}}}{3 \log\left(a/\bar{a}\right)},
	\end{equation}
	where $\bar{a}$ is a constant related to the comoving entropy in the hidden sector. 
Likewise, the yield of dark matter displays the same logarithmic dependence on the scale factor. In the nonrelativistic limit, entropy conservation leads to the relation
	\begin{equation}\label{Y2}
	Y_\mathrm{ent}  = \frac{T_{\rm{D}}}{\xi (m_{\rm{D}} + 5/2 T_{\rm{D}})}\, ,
	\end{equation}
	where $\xi = S/S_{\rm{D}}$ is the entropy ratio.

	In order for entropy to be conserved the temperature has to follow the differential equation
	\begin{equation}\label{T2}
	\frac{d T_{\rm{D}}}{dx} = \frac{2 T_{\rm{D}}^2 (2 m_{\rm{D}}+5 T_{\rm{D}}) \frac{d s}{dx}}{\left(4 m_{\rm{D}}^2+12 m_{\rm{D}} T_{\rm{D}}+15 T_{\rm{D}}^2\right) s}.
	\end{equation}
	In the case where $T_{\rm{D}}\ll m_{\rm{D}}$, we see that the temperature depends logarithmically on the scale factor, since $s^\prime(x)/s = -3a^\prime(x)/a$, as shown in \cite{Carlson:1992fn}.

\begin{figure}[t]
		%\centering
		\includegraphics[width=0.70\columnwidth]{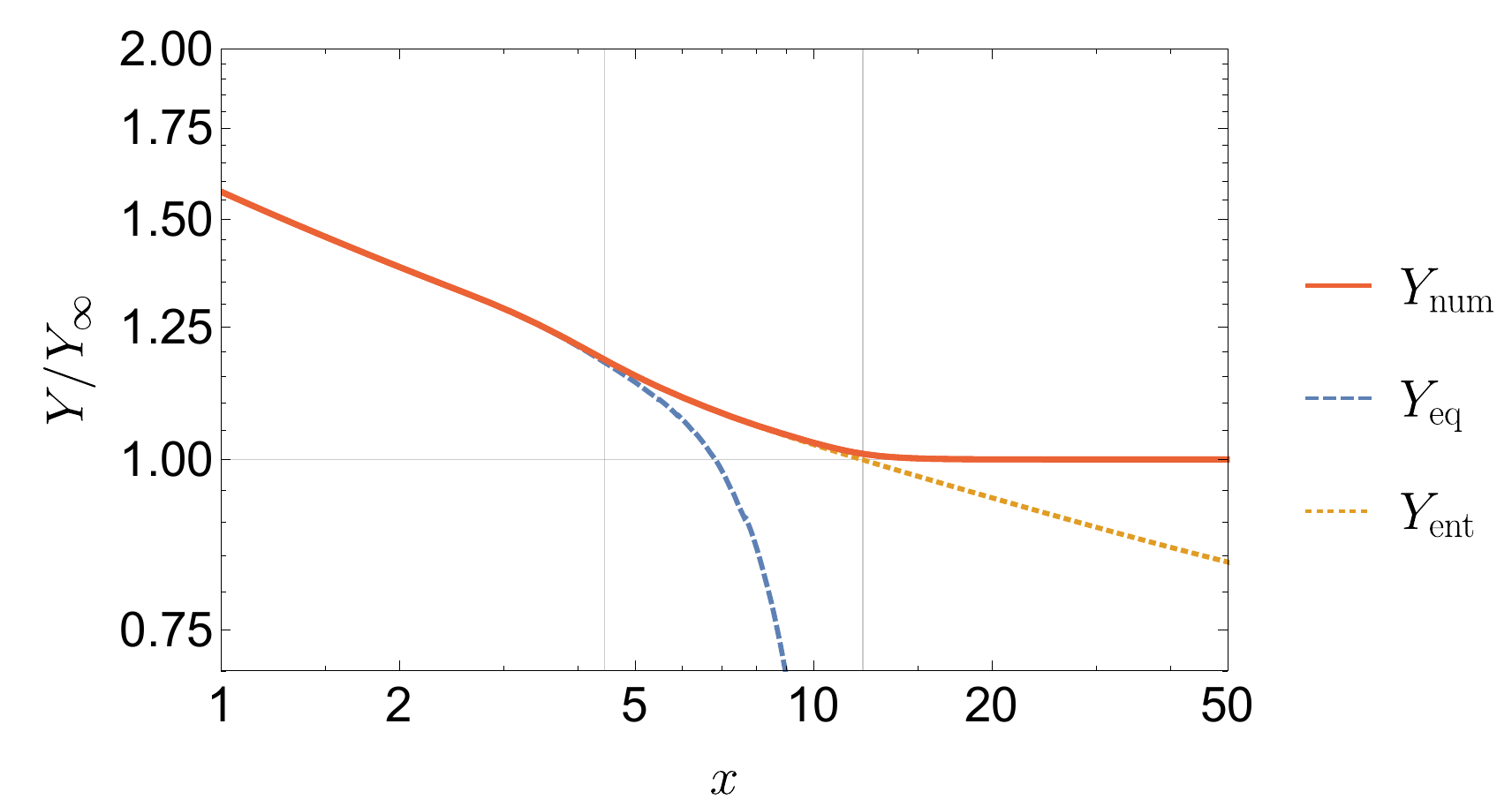}
		\includegraphics[width=0.70\columnwidth]{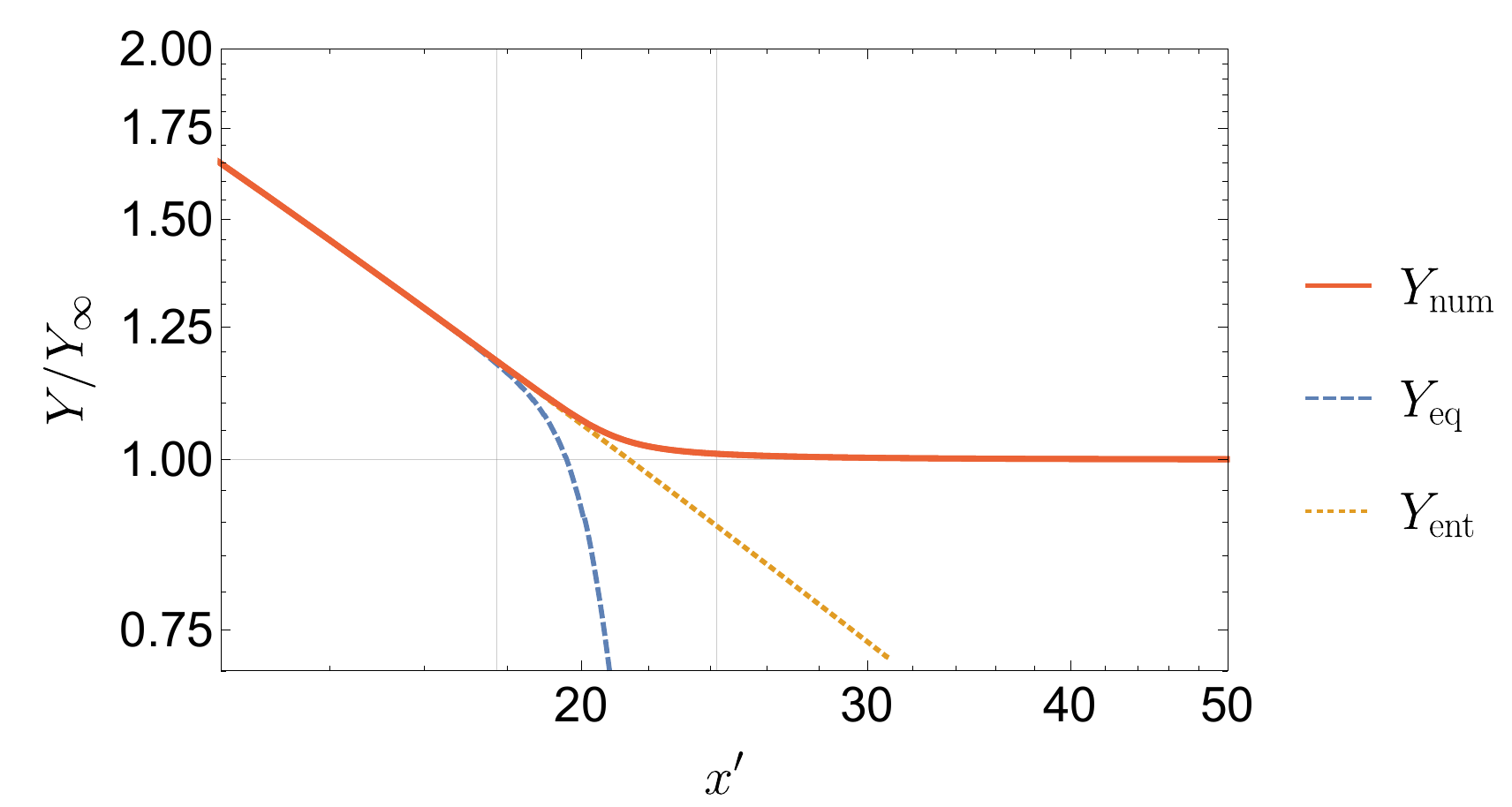}
		\caption{The numerical solution of the yield $Y_\mathrm{num}$ (solid red) together with the equilibrium result $Y_\mathrm{eq}$ (dashed blue) and the entropy-conservation result $Y_\mathrm{ent}$ (dotted orange). The yields are normalized to the observed relic yield.   The results are functions of $x = m_{\rm{D}} /T$ ($x' = m_{\rm{D}} /T_{\rm{D}}$) in the upper (lower) panel. The first vertical line is the point $x_0$ where $T_{\rm{D}}$ starts deviating from Eq.~\eqref{T2}, while the second line marks the freeze-out temperature. The benchmark is for $m_{\rm{D}} = 1\, \mathrm{GeV}$ and with the temperature ratio at freeze-out $T_{\rm{D}}/T = 1/2$. Note that the mapping from $x$ to $x'$ is not the same for the three yields after the first vertical line. }
		\label{fig:SIMP1}
	\end{figure}

\begin{figure}[t]
		\centering
		\includegraphics[width=0.70\columnwidth]{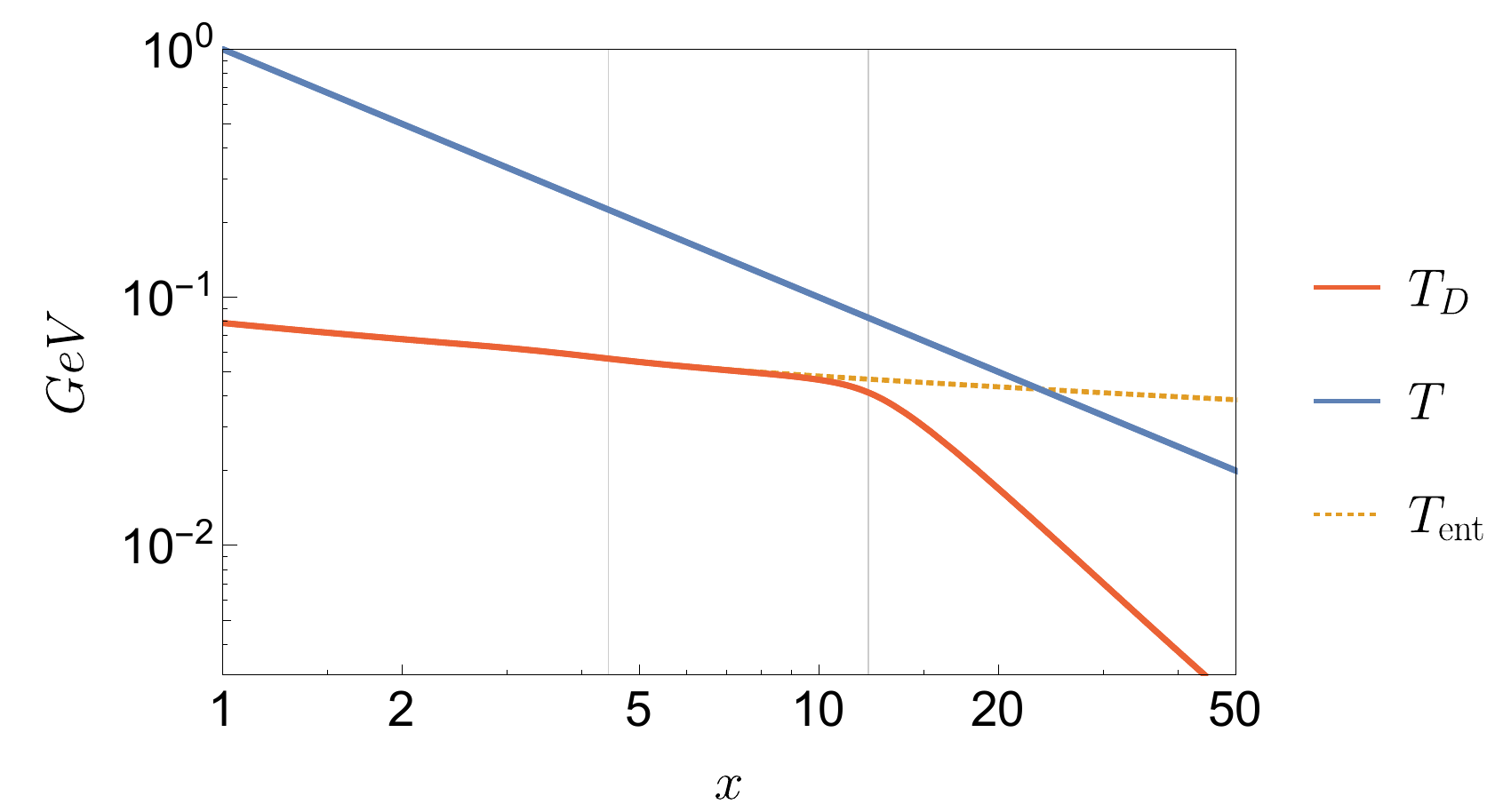}
		\includegraphics[width=0.70\columnwidth]{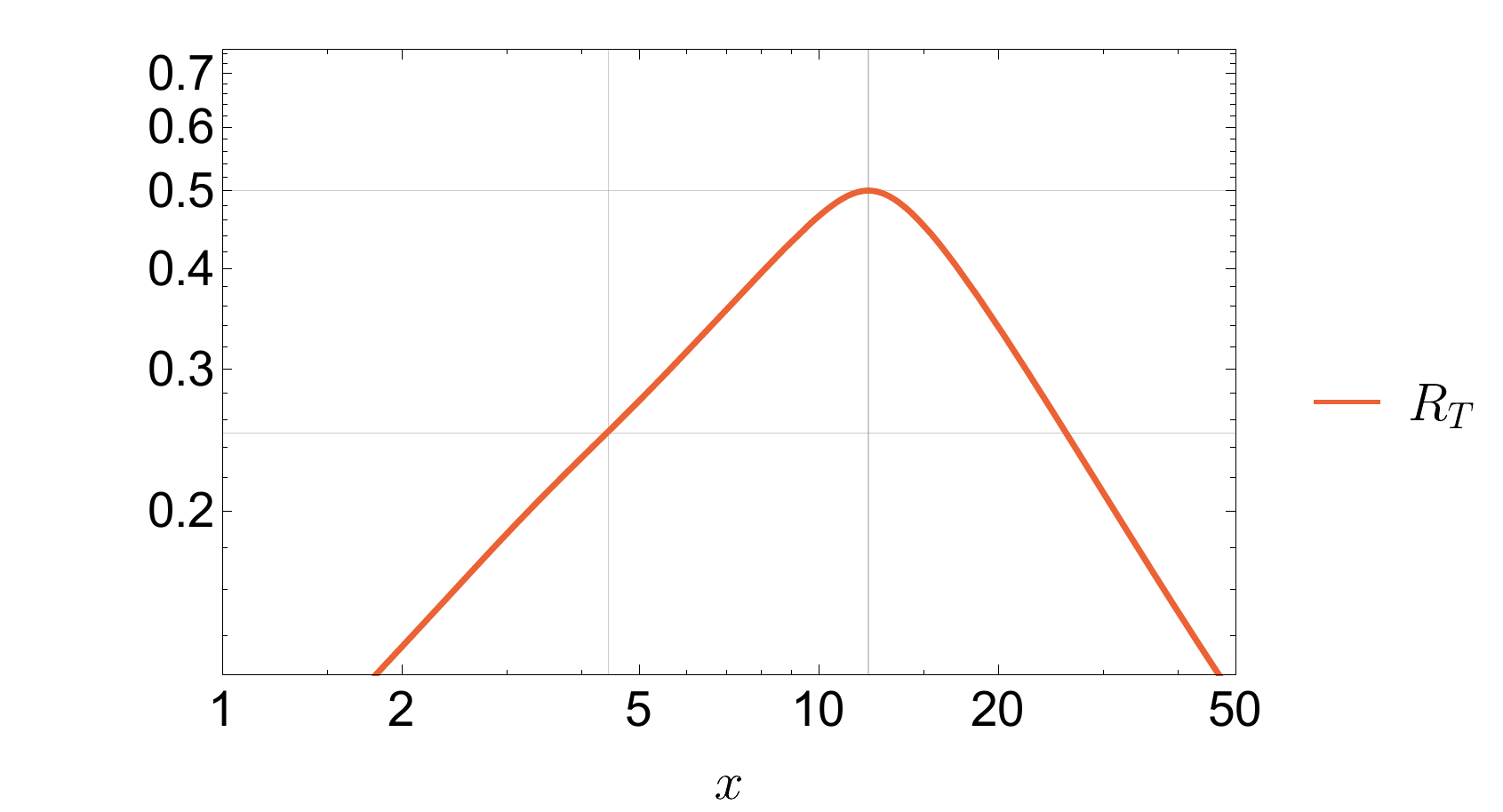}
		\caption{{\it Top panel}: The $x$ dependence of $T$ (blue) and $T_{\rm{D}}$ (red) together with the $T_{\rm{D}}$ solution (dotted orange) enforcing entropy conservation in the dark sector. {\it Bottom panel}: The temperature dependence of the ratio $R_T = T_{\rm{D}}/T$. The first vertical line is the point $x_0$ where $T_{\rm{D}}$ starts deviating from Eq.~(\ref{T2}), while the second line marks the freeze-out. The horizontal lines mark the temperature ratios at $x_0$ and $x_{\rm{FO}}$. The benchmark is for $m_{\rm{D}} = 1\, \mathrm{GeV}$ and with the temperature ratio at freeze-out $T_{\rm{D}}/T = 1/2$.}
		\label{fig:SIMP2}
	\end{figure}

	On the other hand, the temperature evolution must be determined by the expansion of the Universe and the conversion of rest energy into kinetic energy when the number of comoving particles is decreased, i.e.
	\begin{equation}\label{BT}
	\frac{d T_{\rm{D}}}{dx} = \frac{2 T_{\rm{D}}}{3}\frac{1}{s}\frac{d s}{dx}-\frac{T_{\rm{D}}}{Y}\frac{d Y}{dx}\left(1+\frac{2}{3}\frac{m_{\rm{D}}}{T_{\rm{D}}}\right).
	\end{equation}
	When $Y$ is given by Eq.~(\ref{Y2}), the above equation reduces to Eq.~(\ref{T2}), whereas when $Y^{-1}dY/dx\simeq 0$ we see that $T_{\rm{D}} \propto a^{-2}$; i.e. after the number-changing interactions decouple, the hidden sector temperature behaves like that of nonrelativistic matter. We solve Eqs.~(\ref{Boltzmann-Y}) and (\ref{BT}) numerically. A benchmark solution is shown in Figs.~\ref{fig:SIMP1} and \ref{fig:SIMP2}.

	From Figs.~\ref{fig:SIMP1} and \ref{fig:SIMP2}, we see that when the number-changing processes are only slightly insufficient, the temperature will decrease faster than logarithmically and the equilibrium number density will start to display its exponential dependence on the scale factor instead of Eq.~(\ref{Y2}). As a result, the chemical potential is nonzero, and the number-changing processes, being primarily one-way, are enhanced.
  This lasts until the point of freeze-out, where the comoving number density is fixed.

	We define $x'_0$, where $Y$ starts deviating from Eq.~(\ref{Y2}), as the point where $dT_{\rm{D}}/dx$ deviates by 1\% from Eq.~(\ref{T2}). This in turn means that $Y^{-1} dY/dx$ roughly deviates with $3/(200 x^\prime)$, which is negligible. However, it indicates that $Y-Y_{\mathrm{eq}}\sim \cO(10^{-2}) Y_{\mathrm{eq}}$. From these conditions and Eq.~(\ref{Boltzmann-Y}), we get
	% % %
	\begin{equation}\label{BY2}
	\Gamma \simeq   \frac{300 H}{3+x'_0}  \, ,
	\end{equation}
	% % %
	which can be compared with the usual freeze-out condition  $\Gamma \sim H$.
	In the following, we estimate freeze-out to happen roughly at $x'_0+3$, and from a linear extrapolation we get that $Y(x_0)=(1+3 x'^{-1}_0)Y(\infty)$ in order to produce the right amount of dark matter. This leads to an estimate for the cross section
	% %
	\begin{align}\label{sigma}
	\langle\sigma v^2\rangle_{3\to2}
	\simeq \frac{300\, H x'^2_0 }{s^2 Y_\infty^2 (x'_0+3)^3}
	\simeq  \left(1293.5\,\mathrm{GeV}^{-5}\right) \left(\frac{1\,\mathrm{ GeV}}{m_{\rm{D}}}\right)^2 g_{\mathrm{eff}}^{1/2}h_{\mathrm{eff}}^{-2}\, x^4_0 x'^{-1}_0.
	\end{align}
	% %
	This has to be solved together with the condition $Y = (1+3 x'^{-1}_0) Y(\infty)$, which roughly corresponds to the implicit expression Eq.~(\ref{xpFO}).
	When comparing Eq.~(\ref{sigma}) with Eq.~(\ref{FreezeSigma}), we find that Eq.~(\ref{sigma}) introduces an $x'$ dependence.

\begin{figure}[t]
  	\centering
  	\includegraphics[width=0.45\columnwidth]{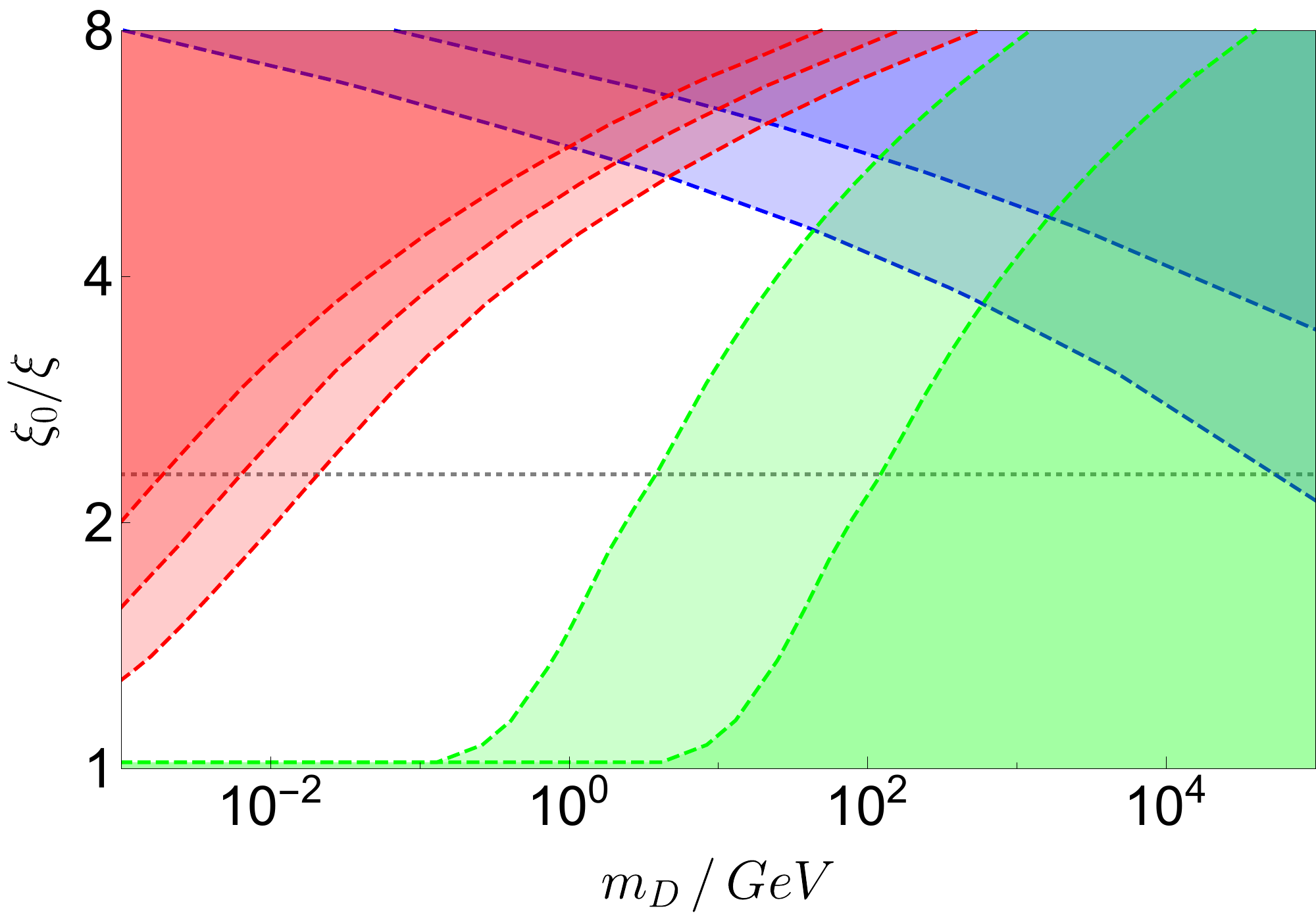}\hspace{0.04\textwidth}
  	\includegraphics[width=0.45\columnwidth]{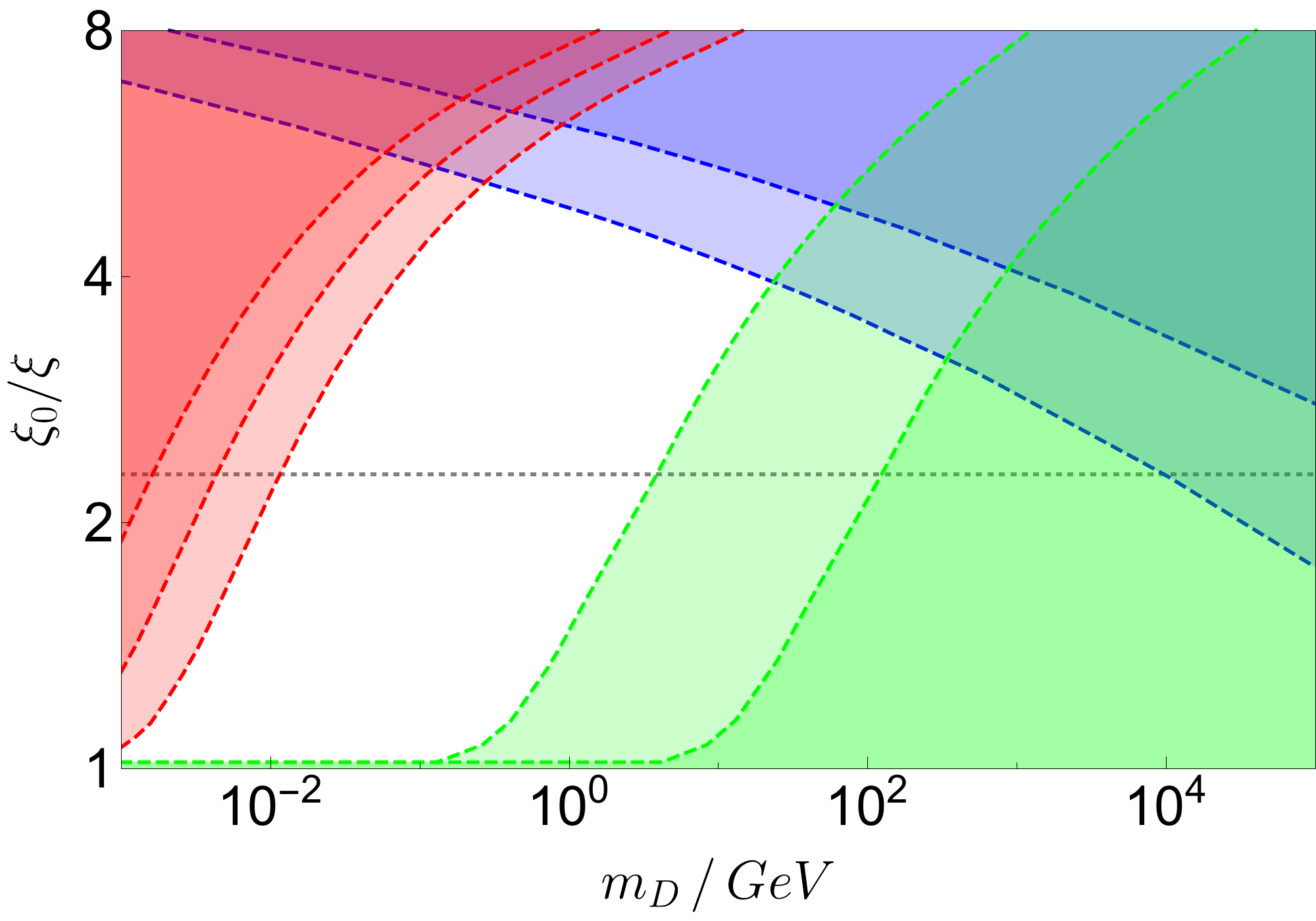}
  	\caption{The parameter space in terms of mass $m$ and initial entropy ratio $\xi_0/\xi$, where $\xi_0$ is the entropy ratio in the ultrarelativistic case, i.e. Eq.~(\ref{xiRatio}). The figure shows a general SIMP model ({\em left}) and
  		the specific SIMP realization ({\em right}) introduced in Sec.~\ref{sec:StandardSIMP}.  The dotted gray line marks the breakdown of the non-relativistic assumption. The blue regions are for $\alpha_{\text{eff}}>1$ and $\alpha_{\text{eff}}>10$ in the left panel and $m_\pi/f_\pi \geq \pi$ and $m_\pi/f_\pi \geq 2 \pi$ in the right panel. The borders of the red regions are for $\sigma/m_{\rm{D}} = \{1, 10^{-1},10^{-2}\}$ in $\mathrm{cm^2/g}$ (left panel assumes $a = 1$). In the green regions $T_\mathrm{FO}>100$~GeV and $T_\mathrm{FO}>10$~TeV. In this plot  $\alpha_{\text{eff}}$ is calculated by assuming the number of DM degrees of freedom is 5, i.e. $g_D =5$.}
  	\label{fig:ent}
  \end{figure}

	Clearly, if we fix the ratio $R_T$ at $x_0$, we can solve the set of Eqs.~(\ref{sigma}) and (\ref{xpFO}).  Since entropy conservation is still valid up until this point, we can replace the input $R_T$ with the entropy ratio $\xi$, and Eq.~(\ref{xpFO}) gives $x'_0 \simeq 18 \,(\xi_0)/(5\xi)$.
From the numerical solution (see Fig.~\ref{fig:SIMP2}), we see that $R_T$ at the point of the estimate is approximately half of the temperature ratio at the time of freeze-out. On the other hand, $\xi$ is nearly constant up until freeze-out. Therefore, we will now investigate the parameter space in terms of $\xi_0/\xi$ and $m_{\text{D}}$, shown in Fig.~\ref{fig:ent}

	Comparing  Eq.~(\ref{Y2}) with the relic density of dark matter today, we see
  that the ratio $m_{\rm{D}}/\xi$ determines the ratio $Y_\mathrm{ent}/Y_\infty$. It
  turns out that if $\xi > \xi_0/2.3$, the temperature already deviates from
  Eq.~(\ref{T2}) for $x'\sim 3$, and freeze-out happens for $x' \sim 5$. We
  define this as the point of breakdown of our nonrelativistic
  assumptions, shown by the gray dotted line in Fig.~\ref{fig:ent}.
  For higher values of $m_{\rm{D}}/\xi$, we need increasingly strong
  interactions to deplete enough dark matter particles to reproduce the
  observed relic abundance. However, at some point the needed interaction
  strength becomes nonperturbative. Therefore, there is a limited range of
  $m_{\rm{D}}/\xi$, where the perturbative description is valid. This constraint is
  illustrated by the blue areas in Fig.~\ref{fig:ent}.
  	% % % %
  %
  Furthermore, light masses will more
  easily give rise to high self-interactions in terms of $\sigma/m_{\rm{D}}$
  and be in violation with the constraints, illustrated by the red areas in
  Fig.~\ref{fig:ent}.

For heavy DM masses, the freeze-out in the hidden sector happens at temperatures which can correspond to SM temperatures significantly above the electroweak phase transition. This is illustrated by the green areas in Fig.~\ref{fig:ent}. While this is not necessarily a problem for the model, it may lead to a more complicated thermal history of the hidden sector than what we have discussed here: Equation (\ref{Boltzmann-Y}) is based on the assumption that the evolution of the hidden sector temperature, and number density is governed solely by the $3\to2$ interaction. Therefore, whatever interaction is responsible for initially creating the hidden sector thermal bath, any energy transfer between the hidden and visible sectors should no longer be present during the thermal evolution leading to freeze-out of the DM abundance, described by the Boltzmann equation (\ref{Boltzmann-Y}). If the hidden and visible sectors are initially coupled through the Higgs portal, this assumption will only hold if the SM temperature during the hidden sector freeze-out is below the electroweak scale. Thus, in the case of an electroweak scale portal, the green areas in Fig.~\ref{fig:ent} do not correspond to a hidden sector freeze-out process, but instead to a reannihilation process as discussed in \cite{Bernal:2015ova}. We will discuss the origin of the hidden sector thermal bath in Sec.~\ref{sec:Discussion}.
Taking all these constraints into account, we see that there is a restricted region in $(m_{\text{D}}, \xi)$-space that fulfills all the criteria.

We will now consider the specific composite realization introduced in Sec.~\ref{sec:StandardSIMP}. Since the $3\to2$ cross section, Eq.~(\ref{eq:crossSIMPlest}), is velocity dependent, the conversion from $\alpha_{\text{eff}}$ to $f_\pi$ will depend on $x'_{\rm{FO}}$. However, in this case we can eliminate the additional coupling $a$. This provides less theoretical uncertainty in placing the self-interaction bounds. At the same time, we can quantify the nonperturbative bounds in terms of the expansion parameter in chiral perturbation theory, namely $m_\pi/f_\pi$.
    
  Let us summarize the three different constraints depicted in the right panel of Fig.~\ref{fig:ent} in terms of the parameters relevant for chiral perturbation theory:
	\begin{enumerate}
		\item {\it Non-perturbative coupling:}\\
		The chiral expansion is a low-energy effective description in which higher order terms are suppressed in terms of the pion mass $m_\pi$ and the pion momentum $p_\pi$ with respect to the scale associated with spontaneous symmetry breaking, $4 \pi f_\pi$. In the right panel of Fig.~\ref{fig:ent}, we mark the regions where the suppression of higher orders is  $m_\pi/f_\pi \geq \pi$ and $m_\pi/f_\pi \geq 2 \pi$, respectively.
		\item {\it Too strong self-interactions:}\\
		Using the expression in Eq.~(\ref{eq:crossSIMPlest}), we enforce the self-interaction bounds. In Fig.~\ref{fig:ent}, we show the lines for $\sigma/m_{\rm{D}} = \{1,10^{-1},10^{-2}\}$ in $\mathrm{cm^2/g}$.
		\item {\it Very early decoupling:}\\
		For heavy dark matter masses and low temperature ratios, the visible sector can still be well above the scale of the electroweak phase transition when freeze-out happens in the hidden sector. We mark the regions corresponding to $T_{\mathrm{FO}} > 100$~GeV and $T_{\mathrm{FO}} >~10$~TeV.
	\end{enumerate}

\begin{figure}[t]
		\centering
		\includegraphics[width=0.70\columnwidth]{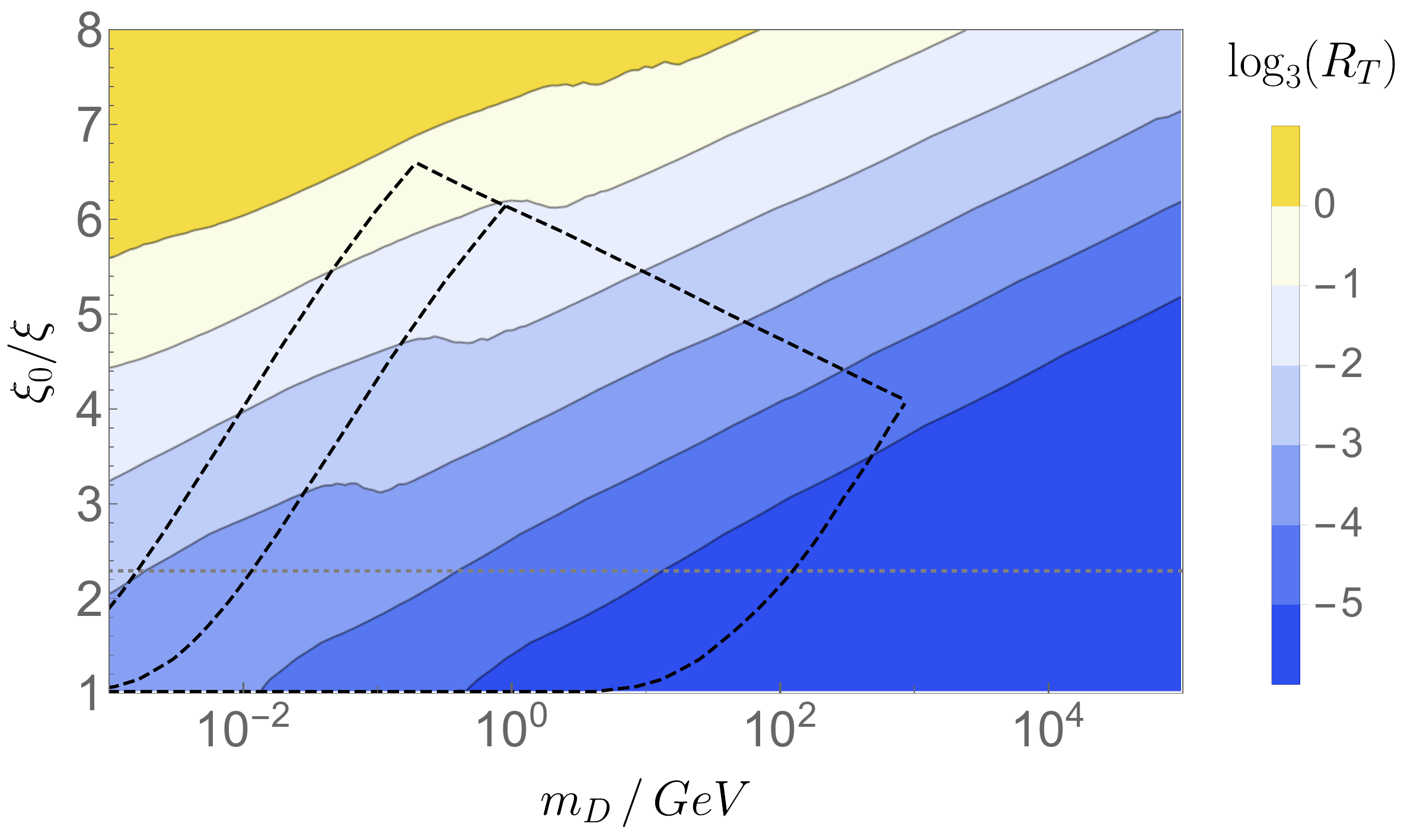}
		\caption{Contours showing the temperature ratio $R_T$ at freeze-out in terms of the dark matter mass $m_{\rm{D}}$ and entropy ratio $\xi$. The contours are for $R_T = 3^n$ with $n$ ranging from $-5$ to 0 (bottom to top). The dashed lines display the constraints illustrated in the right panel of Fig.~\ref{fig:ent}. The gray dotted horizontal line marks the breakdown of the nonrelativistic assumption.}
		\label{fig:TempRatio-SIMP}
	\end{figure}

	Our analysis of the model introduced in Sec.~\ref{sec:StandardSIMP} shows that for the pions to have self-interactions in the range $\sigma/m_{\rm{D}} \in \{1,10^{-2}\}$~$\mathrm{cm^2/g}$, and for chiral perturbation theory to be applicable, the dark matter mass has to be below 1~GeV. In Fig.~\ref{fig:TempRatio-SIMP}, we superpose the constraints from the right panel of Fig.~\ref{fig:ent} on top of contours displaying the temperature ratio $R_T$ at freeze-out. We see that when $R_T\geq 1$, the self-interactions are not in the range $\sigma/m_{\rm{D}} \in \{1,10^{-2}\}$~$\mathrm{cm^2/g}$ while $m_\pi/f_\pi \leq 2 \pi$. For lower values of $R_T$, i.e. $0<R_T<1/3$, some of the favored range of self-interactions is accessible within chiral perturbation theory. For $R_T<1/3$, the whole favored range corresponds to $m_\pi/f_\pi \leq 2 \pi$.
	In the appendix, we show how the model is affected by higher order contributions in the chiral expansion. We find for $R_T<1/3$ that the model is still phenomenologically viable when parametrizing the higher order effects.

\section{Origin of entropy difference and testability}
\label{sec:Discussion}

Generally, a hidden sector at temperature $T_{\rm{D}}$ different than the temperature
$T_{\rm{SM}}$ of the SM heat bath can be created in two different ways: First, it is possible to create the quanta in the hidden sector directly from the inflaton field at reheating~\cite{Adshead:2016xxj}. Second, the hidden sector particles could be produced at lower energies from the SM heat bath via the freeze-in mechanism.

A generic feature of decoupled hidden sectors is that they may lead to
isocurvature fluctuations, which are, on the basis of Planck data, known to
be heavily suppressed. If the matter quanta of the hidden sector originate
from the same source as the visible sector ones, i.e. from the
single inflaton field either directly or via the SM model fields, it is
known that they will inherit the same adiabatic fluctuations~\cite{Weinberg:2004kr}.
The observable isocurvature fluctuations will, however, arise if a primordial
scalar condensate exists in the hidden sector \cite{Nurmi:2015ema,Kainulainen:2016vzv}, and this
will provide nontrivial constraints between the masses and couplings.

Let us consider the freeze-in production of a hidden sector particle species, followed by thermalization within the hidden sector due to number-changing scattering processes. This mechanism has been discussed in the context of Higgs portal models in e.g. \cite{Chu:2011be,Bernal:2015ova,Bernal:2015xba,Heikinheimo:2016yds}.

Here we are interested in a hidden sector model that exhibits a global symmetry breaking pattern such as $SU(2N_f)\rightarrow Sp(2N_f)$, as is realized in models of composite DM. We will therefore outline the creation of a thermal bath of composite DM particles with $T_{\rm{D}}\neq T$. Let us thus write down the Lagrangian for a hidden sector that contains $N_f$ SM-singlet fermions, charged under a hidden sector $SU(2)$ gauge interaction, and a scalar mediator particle that is singlet under both the SM and the hidden sector gauge groups:
\beq
\mathcal{L_D} = -\frac14 F_{D\mu\nu}F_D^{\mu\nu}+\partial_\mu s\partial^\mu s-\sum_f\bar{\psi}_f(m_f+y_f s-i\slashed{D})\psi_f-V(s),
\eeq
where $\psi_f$ are the hidden sector fermions, $F_D$ is the field strength tensor of the hidden sector gauge field and $s$ is the mediator scalar, with Yukawa couplings $y_f$ to the hidden fermions and a potential given by
\beq
V(s) = \frac12\lambda_{sH}s^2H^\dagger H+\lambda_s s^4+\mu_s^2 s^2,
\eeq
where $H$ is the SM Higgs doublet. The portal coupling $\lambda_{sH}$ is the only gauge invariant renormalizable interaction between the hidden sector and SM particles, and it controls the initial freeze-in production of the hidden sector degrees of freedom.

For simplicity, we will for now assume that the confinement scale of the hidden sector gauge theory is somewhere above the electroweak scale, so that for the energy range of interest, the hidden sector is described by the chiral effective theory for the composite pions.\footnote{For the SIMP realization discussed in Sec.~\ref{sec:DarkFreezeOut}, this is in fact not always the case, unless the chiral symmetry breaking scale $\Lambda_{\rm D}$ is very large compared to the pion decay constant $f_\pi$. For $\Lambda_{\rm D}\sim 4\pi f_\pi$ this assumption only holds for dark matter masses above a few GeV. We will remark on this possibility towards the end of this section.} As the hidden sector becomes confined, the  nonzero vacuum expectation value $\langle\bar{\psi}_f\psi_f\rangle$ constitutes a linear term in the scalar potential due to the Yukawa coupling $y_fs\bar{\psi_f}\psi$, resulting in a vacuum expectation value for the field $s$ \cite{Hur:2011sv,Heikinheimo:2013fta,Heikinheimo:2014xza}. Below the electroweak symmetry breaking scale, the portal coupling $\lambda_{sH}$ leads to mixing between the $s$ and $h$ fields, and thus the dark sector degrees of freedom can be produced via Higgs decays, assuming $m_{\pi}<\frac12 m_H$, where $m_{\pi}$ is the mass of the composite DM particle. Then, the number density of composite DM produced via Higgs decays is approximately \cite{Chu:2011be}
\beq
n_D^{\rm initial} = \left. 3\frac{n_h^{\rm eq}\Gamma_{h\to\pi_D\pi_D}}{H}\right|_{T=m_h},
\eeq
where $n_H^{\rm eq}$ is the equilibrium number density of Higgs bosons in the SM plasma. Below $T\sim \frac13 m_h$ the Higgs decouples from the SM plasma, and the energy transfer between the hidden and the visible sector stops so that entropy densities within both sectors are conserved separately. The initial energy density of the hidden sector is given by $\rho_D^{\rm initial}=\frac 12 m_h n_D^{\rm initial}$, and assuming instant thermalization of the hidden sector, the initial temperature of the hidden sector is then
\beq
{T}_D^{\rm initial} = \left(\frac{30 \rho_D^{\rm initial}}{g_*^D\pi^2}\right)^\frac14,
\label{eq:Tinit}
\eeq
where $g_*^D$ is the number of light degrees of freedom in the hidden sector (the number of pions). The conserved hidden sector entropy density is thus
\beq
S_{\rm{D}} = \frac{g_*^D 2\pi^2}{45}({T}_D^{\rm initial})^3.
\label{eq:Sinit}
\eeq

This would then provide the initial condition for the analysis
carried out in Sec.~\ref{sec:DarkFreezeOut}.

As we noted above, the simplifying assumption that chiral effective theory is a valid description of the dynamics of the hidden sector throughout its thermal history might not always hold, depending on the ratio of the symmetry breaking scale to the pion decay constant. If this is not the case, the initial production of the hidden sector particles should be described by the degrees of freedom of the unbroken phase, the fermions $\psi_f$ and the gauge fields of the confining gauge group. However, as long as the phase transition happens at a scale well separated from the scale of production of the hidden sector thermal bath, and of the eventual freeze-out of the composite degrees of freedom, the dynamics of the phase transition should not affect the overall picture very much. As long as the entropy of the hidden sector is approximately conserved in the phase transition, Eqs.~(\ref{eq:Tinit}) and (\ref{eq:Sinit}), relating the temperature and entropy density of the hidden sector to the initially produced energy density, will give a correct description that can be used as an input for our analysis presented in Sec.~\ref{sec:DarkFreezeOut}.

Generally, at a first order finite temperature phase transition in the early Universe, gravitational waves are generated~\cite{Witten:1984rs,Hogan:1986qda,Kosowsky:1992rz,Hindmarsh:2013xza}. Such a transition in the hidden sector would therefore provide an indirect signal for the
hidden sector dynamics~\cite{Schwaller:2015tja,Tsumura:2017knk,Aoki:2017aws}. The numerical results of the previous section predict, using the naive estimate $\Lambda_{\rm D} \sim 4 \pi f_\pi$, that in the region with favorable self-interactions, the hidden sector will go through the chiral phase transition when the temperature of the visible sector is below 100 GeV. However, determining the order and the detailed dynamics of the chiral phase transition is beyond the scope of this work.

Finally, we note that dark matter feebly interacting with the standard
model and thus inaccessible for collider or direct detection searches, could still be detected indirectly via annihilations, e.g. at the Galactic center. Due to the smaller hidden sector temperature at the time of DM freeze-out, this mechanism results in a different relation between the DM annihilation rate during freeze-out and today, as compared to the standard freeze-out scenario with $T_{\rm{D}}=T$, and generally the expected amplitude of the indirect detection signal is weaker~\cite{Heikinheimo:2018duk}. In the case where the DM abundance is determined by the $3\rightarrow2$ process, as discussed in this work, the annihilation signal could be completely absent, although this depends on the details of the hidden sector. Here we have assumed that all the pion species that constitute the low-energy degrees of freedom of the hidden sector are stable, and thus no visible signal is created from pion-pion scattering. However, if some unstable species are present in the low-energy theory, an indirect detection signal could be generated via $2\rightarrow 2$ scattering within the hidden sector, where stable DM pions scatter into the unstable species, which then decay into visible channels~\cite{Heikinheimo:2014xza}.

\section{Conclusions}\label{sec:Conclusion}

In this paper we have considered SIMP
dark matter. We have extended the simplest SIMP realizations studied in the
literature by assuming that the SIMP resides in a hidden sector feebly
coupled with the SM.

We showed that such a dark sector, independently of how it was created, can
reach an internal chemical equilibrium due to number-changing $2\leftrightarrow 3$ processes, at a temperature
distinct from that of the SM thermal bath. The dark matter abundance, then, is determined by the dark freeze-out, i.e. the decoupling of the $3\rightarrow 2$ process. We showed that the observed dark
matter abundance can be achieved and, in the case of a composite SIMP, the analysis can be consistently carried out within chiral perturbation theory.

As a concrete model building framework we outlined how the hidden sector can be
populated by the freeze-in mechanism from the particles in the SM heat bath. Alternatively, in the absence of any direct coupling between the SM and the hidden sector, the SIMP degrees of freedom could be produced directly from the decay of the inflaton field at reheating.

The hidden sector SIMP, with its own thermal history, therefore
provides an attractive model building framework for self-interacting dark matter, and allows for a controlled perturbative treatment within chiral perturbation theory.

\section*{Acknowledgements}
This work has been supported by the Academy of Finland, Grant NO. 310130. The work of K.L. is partially supported by the Danish National Research Foundation, Grant No. DNRF90.

% % % % % % % % % % % % %
% % % % % % % % % % % % %
\newpage
% % % % % % % % % % % % %
% % % % % % % % % % % % %
\appendix
\section*{Appendix: Higher order chiral perturbation theory}
\begin{figure}[t]
	\centering
	\includegraphics[width=0.70\columnwidth]{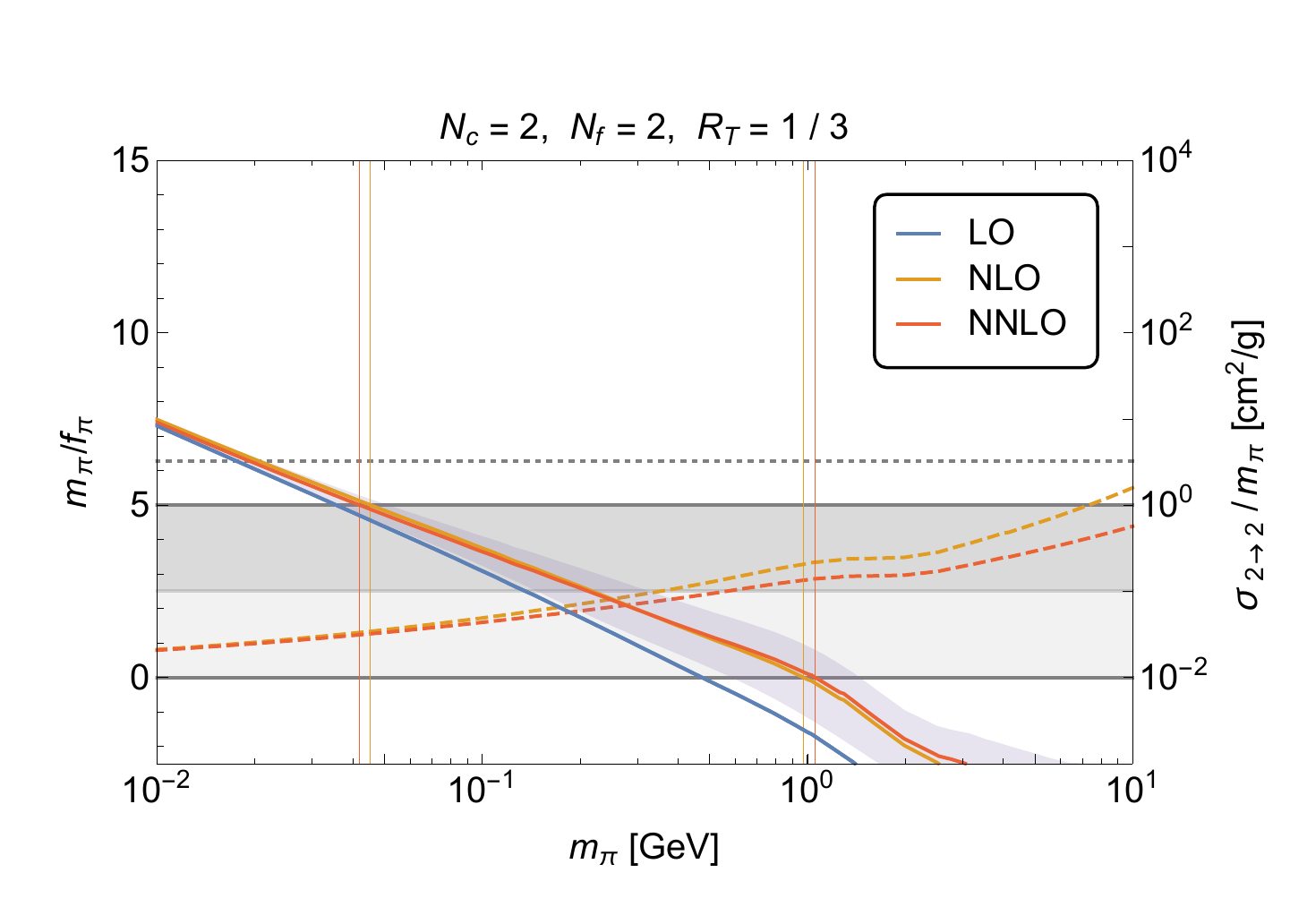}
	\includegraphics[width=0.70\columnwidth]{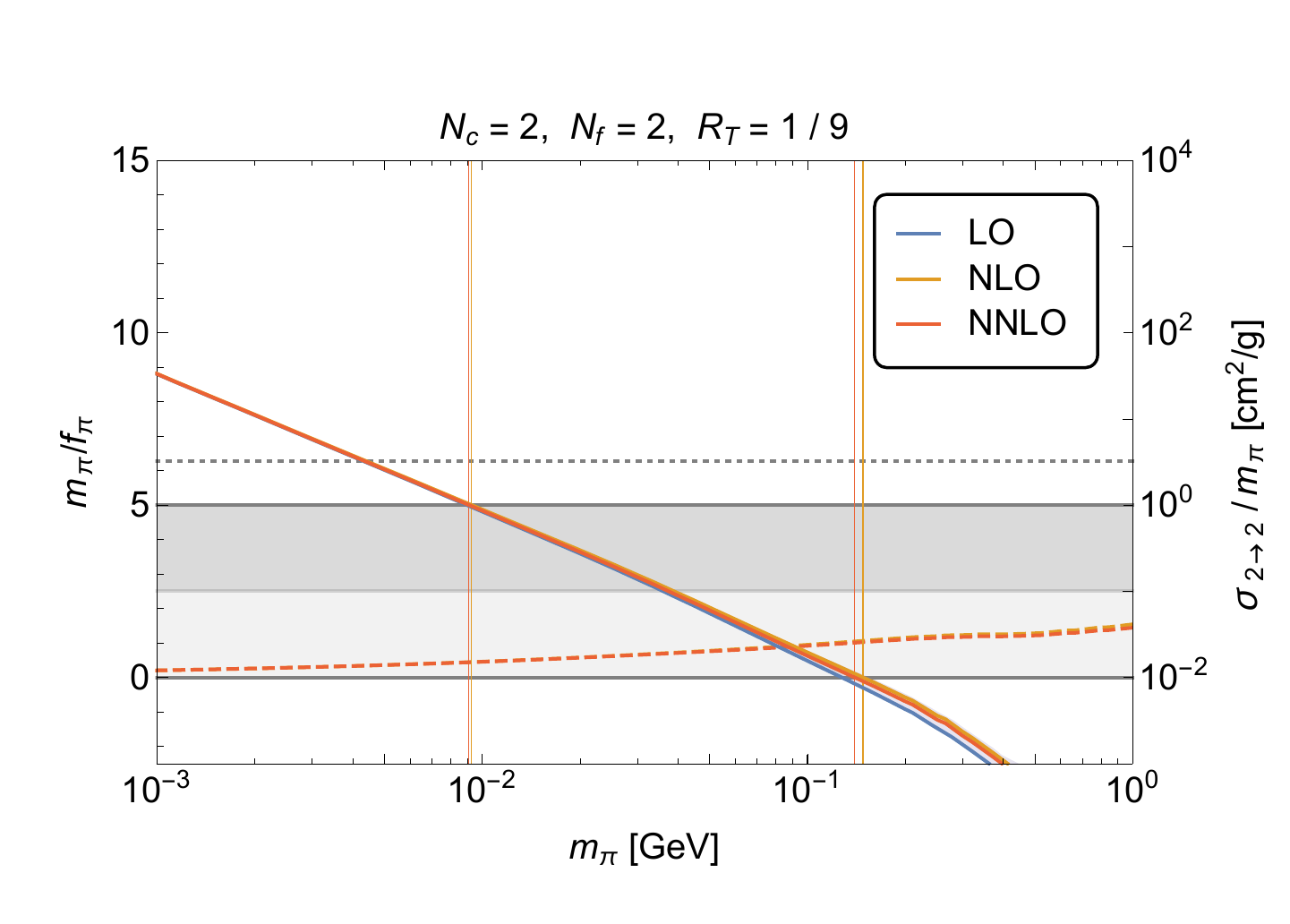}
	\caption{\textit{Dashed} lines are read off on the left axis and \textit{solid} lines are read off on the right axis. The red dashed line is the NNLO solution $m_\pi/f_\pi$ to the Boltzmann equation, the orange dashed is the NLO, and the dashed (grey) horizontal line is the upper perturbative limit $m_\pi/f_\pi=2\pi$. The three solid lines are the cross sections for the $2\to2$ self-interactions at LO (blue), NLO (orange) and NNLO (red). The purple band is the uncertainty from the low-energy constants. The solid grey band is the favored range in self-interaction $\sigma/m_\pi = \{10, 1, 0.1\}$ in $\mathrm{cm^2/g}$. Vertical lines mark the points where the $2\to2$ cross section $\sigma/m_\pi $, at each order in chiral perturbation theory, is entering and exiting the gray band.}
	\label{fig:SIMP}
\end{figure}

The strength of the composite SIMP models is that their strongly coupled fundamental degrees of freedom, at low energies, can be parametrized using only two low-energy constants ($m_\pi$ and $f_\pi$) at lowest order in the chiral expansion. However, when the ratio $m_\pi/f_\pi > 2 \pi$, the model becomes sensitive to an exponentially fast growing number of undetermined low-energy constants from higher orders. It was shown in \cite{Hansen:2015yaa} that, in general, these higher order terms render the minimal choices phenomenologically inviable.
In this work, we have presented a scenario that leaves the minimal models more amenable to the chiral perturbation theory. In this section, we will illustrate the effect of higher order corrections on the most minimal case: an Sp($N_c$) gauge theory ($N_c$ even) with four Weyl fermions in the fundamental $N_c$-dimensional representation, which follows the chiral symmetry breaking pattern $\text{SU}(4)\to\text{Sp}(4)$. Concretely, we single out the cases with $R_T = 1/3$ and $R_T = 1/9$. For further details on how to estimate the size of these higher order terms, we refer to \cite{Hansen:2015yaa}. The results are shown in Fig.~\ref{fig:SIMP}.
%
% % % %

From Fig.~\ref{fig:SIMP} we note that there is a significant correction from NLO. In \cite{Hansen:2015yaa} it is argued that the LO is a mixed order and NLO is the lowest order that correctly describes the model. From the top panel ($R_T = 1/3$), we see that the mass range corresponding to the self-interaction range $\sigma/m_\pi = \{10, 1, 0.1\}$ in $\mathrm{cm^2/g}$ at NNLO is affected by the low-energy constants. However, the averaged result gives a decent estimate, and the viability of the model is not expected to be invalidated by higher order terms. The predicted mass range is from 40~MeV to around 1~GeV, where the upper end has uncertainties from higher order corrections of the order of 500~MeV. The effects from higher order terms are much less notable on the lower panel ($R_T = 1/9$), where the predicted mass range is from 9~MeV to 140~MeV with an uncertainty from higher order corrections of the order of 10~MeV. For lower values of $R_T$, we find a shift towards lower dark matter masses.

\end{document}